\documentstyle[12pt,bezier,epsf,epsfig]{article}
\def\ba{\begin{eqnarray}}
\def\ea{\end{eqnarray}}
\begin{document}
 \begin{flushleft}
 DESY 97--167
 \\
 hep-ph/9709207
 \\
 September 1997
 \end{flushleft}
 
 \noindent
 \vspace*{0.50cm}
\begin{center}
 \vspace*{2.cm} 
{\huge  
Semi-analytic predictions for 
\vspace*{3mm}
\\
$W$ pair production at 500 GeV
 \vspace*{2.cm}               
}
\\
{\large 
J. Biebel and T. Riemann
}
\vspace*{0.5cm}
 
\begin{normalsize}
{\it
Deutsches Elektronen-Synchrotron DESY, Institut f\"ur Hochenergiephysik
\\ 
IfH Zeuthen, Platanenallee 6, D-15738 Zeuthen, Germany
}
\end{normalsize}
\end{center}
 
 \vspace*{1.5cm} 
 \vfill 

\begin{abstract}
We give a short review on the semi-analytic approach to off-shell $W$
pair production at 500 GeV and above.
Numerical calculations are performed with the Fortran program {\tt
  GENTLE} v. 2.0 and its upgrade.
Recent extensions include: 
\\
(i) angular distribution for the {\tt CC11} process with LLA QED
corrections; 
\\
(ii) anomalous couplings.
\\
As an application we study the sensitivity of total cross-section
and forward-backward asymmetry to anomalous couplings.
A noticeable sensitivity from the asymmetry appears only 
for a parity-violating anomalous coupling.
\end{abstract}  

 \normalsize
 \vfill
 \vspace*{.5cm}
  
 \bigskip
 \vfill
 \footnoterule
 \noindent
 {\small
 Emails: {\tt biebel@ifh.de}, {\tt riemann@ifh.de} 
 }
 \newpage
$W$ pair production in $e^+e^-$ annihilation at high energies
has to be studied as off-shell production with at least four fermions in
the final state
(\cite{Altarelli:1996ab,Beenakker:1996ka,Accomando:1997wt} and 
references therein).
A convenient method for this is Monte Carlo simulation. 
In several respects, the semi-analytic approach of the Fortran
program {\tt GENTLE}\footnote{
A detailed technical documentation of  {\tt GENTLE}, version 2.0, may
be found in 
\cite{Bardin:1996zz}. The approach is described 
in \cite{Bardin:1994} (general features),
\cite{Bardin:1993a,Bardin:1996aa} ($W$ pairs),
\cite{Bardin:1995vm,Bardin:1996jw} ($ZZ$ pairs),
\cite{Bardin:1995xx} ($ZH$ production),
\cite{Leike:1995,Biebel:1997aa} (angular distributions, anomalous
couplings).  
}
may prove useful:
\begin{itemize}
\item   Tests and comparisons of different numerical programs
  \cite{Bardin:1996?1,Boudjema:1996qg}; 
\item   Scan of $W$ pair excitation curve \cite{Kunszt:1996km,
    Barate:1997bf,Acciarri:1997ra};
\item   Search for anomalous triple boson couplings
  \cite{Gounaris:1996rz,Acciarri:1997xc}. 
\end{itemize}
The latter topic is better studied by multi-differential
distributions/helicity analysis with Monte Carlo programs
\cite{Gounaris:1996rz,Bardin:1996?1}; but with a
limited luminosity a single differential distribution, derived in a
semi-analytical approach,  may be a reasonable alternative.

The prospects of $WW$ physics at a linear collider with center-of-mass
energy of 500 GeV or more have been summarized in section 5.1 of
\cite{Accomando:1997wt}.
Here, we study as an example the discriminative power of
$W$ pair production 
with respect to parity violating and conserving
anomalous triple boson couplings.
We assume $\sqrt{s} = 500$ GeV and an integrated luminosity ${\cal L} =
50$ fb$^{-1}$.
This corresponds to about 80~000 semi-leptonic $W$ pair decays.
We make a simple ansatz.
Let us explain it with only one anomalous coupling $A$.
For this case, the cross-section may be written as follows: 
\ba
\sigma_{theor}^{(1)} &=& \sigma_{SM} + A  \sigma_1 + A^2 \sigma_{11}.
\ea
The anomalous coupling appears in the cross-section at most
bilinearly.
After having calculated $\sigma_{SM}, \sigma_1, \sigma_{11}$ with {\tt
  GENTLE} (or another program) within a given model and for fixed
experimental conditions, one may confront an experimental data point  
with the predictions and get, in general, two possible solutions for $A$
(or none). 
In a next step, one may consider two anomalous couplings $A$ and $B$
at once, etc.:
\ba
\sigma_{theor}^{(2)} &=& \sigma_{SM} + A  \sigma_1 + A^2 \sigma_{11}
      +B \sigma_{2} + B^2 \sigma_{22} +AB \sigma_{12}.
\label{2fl}
\ea
For our study of sensitivities, we use $\sigma_{theor}^{(2)}$ for the
simulation of 
$\sigma_{meas} \pm \sqrt{\sigma_{meas}/(6{\cal L})}$, the assumed
measured  
cross-section with $1\sigma$ deviations of the counting rates for the
sum of all semi-leptonic production channels. 
For definiteness, we use for $\sigma_{meas}$ the standard model
prediction and apply no experimental cuts. 
The solutions of eq.~(\ref{2fl}) are ellipses in the
plane spanned by the two anomalous couplings.
Allowed pairs of coupling values are located in the area between
the two limiting ellipses.

For the analysis, we use $\sigma_{F}$ and $\sigma_B$, the forward
and backward cross-sections. 
The forward (backward) cross-section is defined by
the requirement that the angle between the momenta of the $e^+$ and
the $W^+$ is less (more) than $90^\circ$. 
We chose these observables since they may be used to form the total
cross-section $\sigma_{tot} = \sigma_{F} + \sigma_B$ (arizing from
cross-section parts even in the production angle) and the forward backward
asymmetry 
$A_{FB} = ( \sigma_{F} - \sigma_B)/\sigma_{tot}$ 
(arising from odd cross-section parts).

\begin{figure}
  \begin{center}
    \epsfxsize=7.3cm
    \leavevmode
    \epsffile{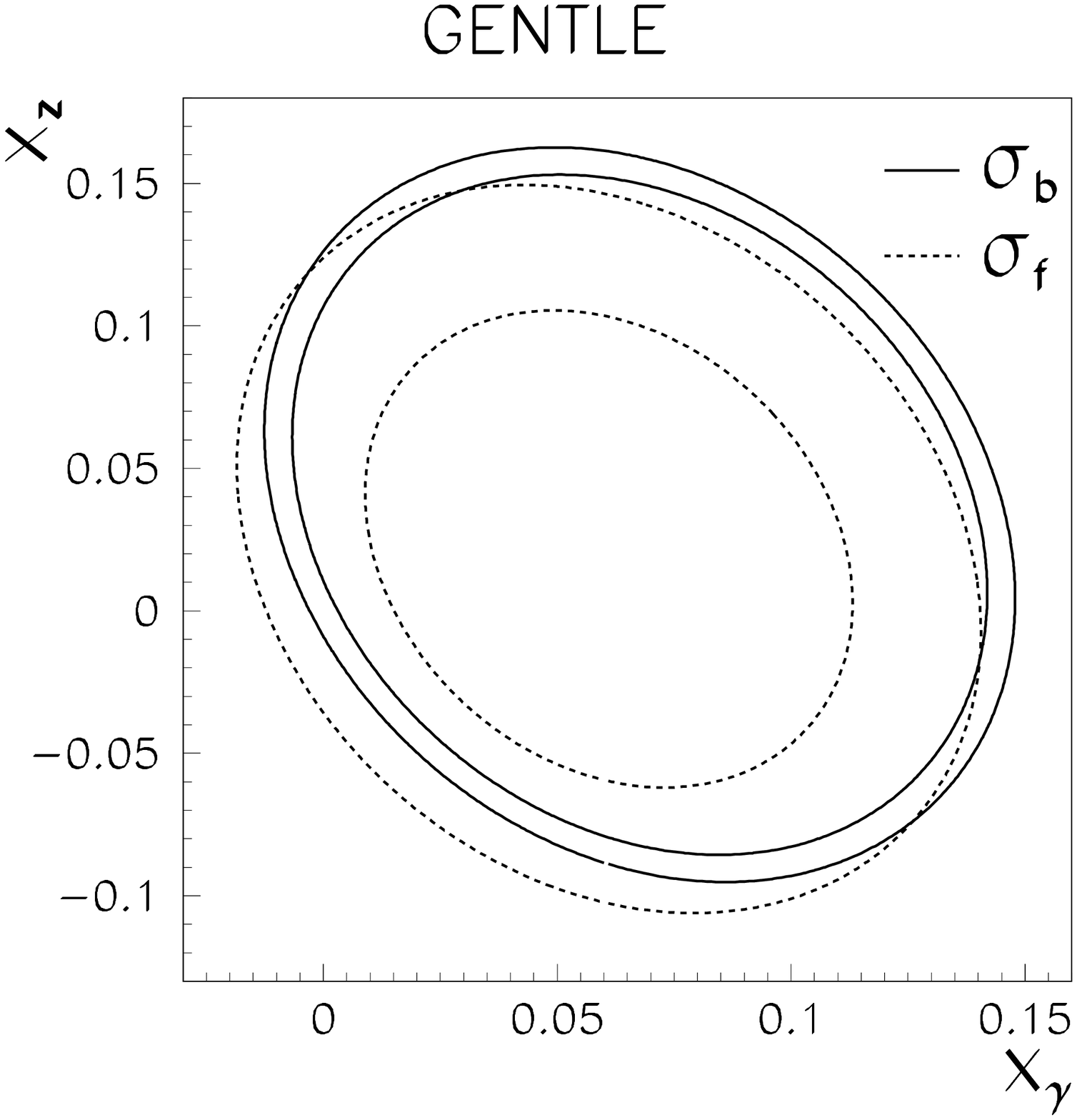}
    \epsfxsize=7.3cm
    \leavevmode
    \epsffile{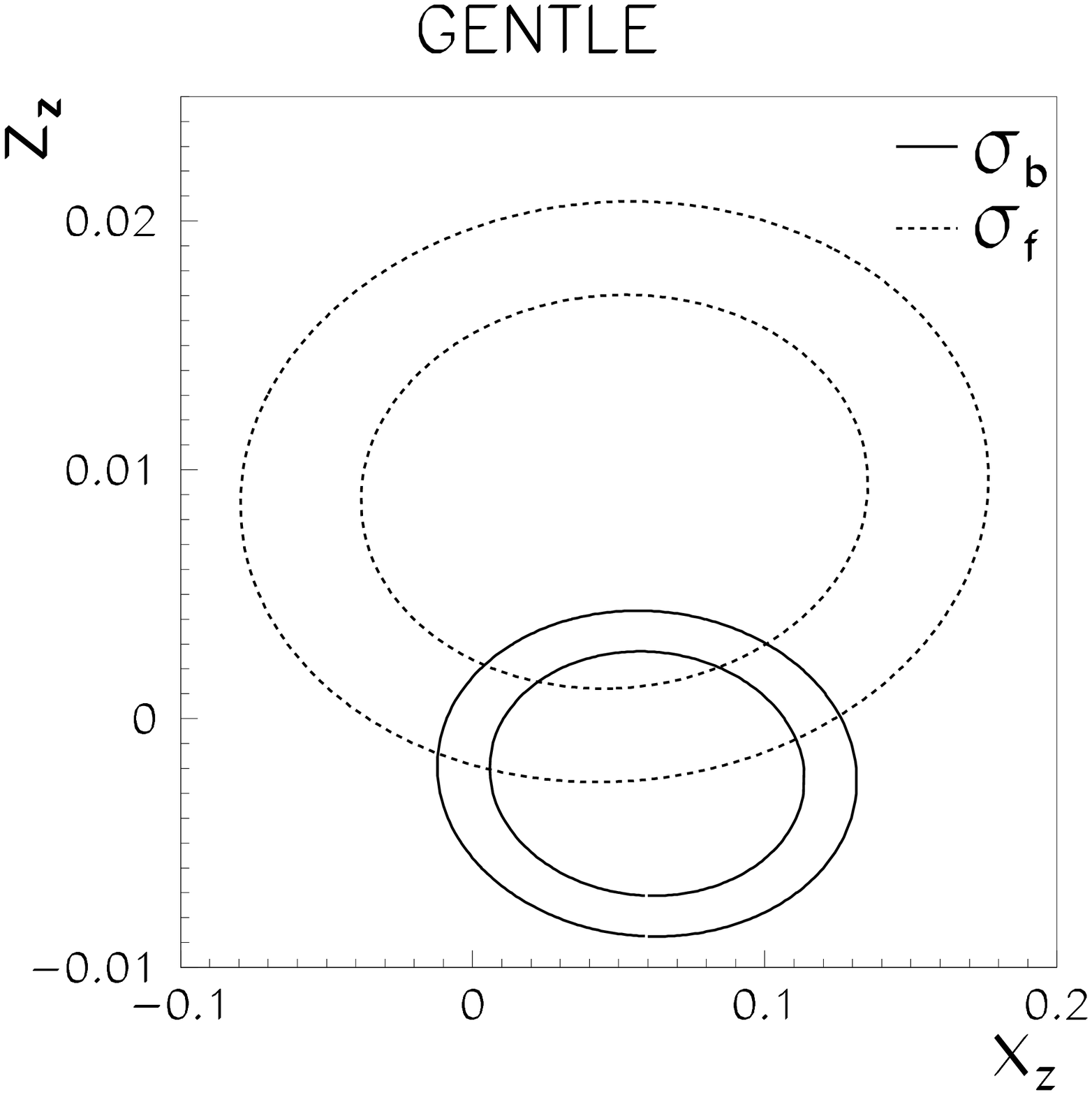}
  \end{center}
\begin{center}
\caption{\label{rings} $1\sigma$-bounds at 500~GeV for ${\cal L}=50 fb^{-1}$.}
\end{center}
\end{figure}

\bigskip

With the definition of the anomalous couplings $x_\gamma$ and $x_Z$ we
follow equation (3) of \cite{Gounaris:1996rz}
which is a re-parameterization of equation (1) of  \cite{Gounaris:1996rz}.
Additionally, we look at the triple boson coupling coupling $z_Z$
defined by the effective Lagrangian 
\begin{eqnarray}
  {\cal L}_Z &=&\mbox{}-\frac{e z_Z}{m^2_W} \,
  \partial_\alpha\hat{Z}_{\rho\sigma}
  \left( W^{+\alpha}\stackrel{\leftrightarrow}{\partial^\rho}W^{-\sigma}-
  W^{+\sigma}\stackrel{\leftrightarrow}{\partial^\rho}W^{-\alpha}\right)~
.
\end{eqnarray}

In figure~\ref{rings}, we show two solutions of eq.~(\ref{2fl}) for
different anomalous couplings. 
In the left figure, $x_\gamma$ and
$x_Z$ are allowed to differ from zero. 
These are ${\cal P}$ conserving couplings and
the two rings derived from $\sigma_{F}$ and $\sigma_B$ overlap almost
totally. 
In the second figure, $x_Z$ and 
$z_Z$ are varied. 
Since $z_Z$ is ${\cal P}$ violating, the allowed ranges overlap much
less since the forward-backward
asymmetry is more sensitive to this coupling.
One further notes that $\sigma_B$ is more
sensitive to anomalous couplings than $\sigma_F$ although the relative
statistical error of the latter is much smaller.

\bigskip

\noindent
{\bf
Acknowledgement
}
\\
We would like to thank the convener of the Subgroup for Study of
Electroweak Gauge Bosons, Thorsten Ohl, for numerous discussions,
hints, and numerical comparisons with the Fortran
program {\tt WOPPER} \cite{Anlauf:1994sc,Anlauf:1996}.

\newpage

\def\href#1#2{#2}
\bibliography{%
lc_jbtr%
/home/phoenix/riemann/Bibliography/zfitter.bib%
}
\bibliographystyle{/home/phoenix/riemann/Bibliography/utphys_tr}
\end{document}